\documentclass[10pt,preprint]{aastex}
\usepackage[utf8]{inputenc}
\usepackage{amsmath}
\usepackage{longtable}
\usepackage{graphicx}

\usepackage{ulem,color}

\title{The K2-TESS Stellar Properties Catalog}
\author{Keivan G.\ Stassun\altaffilmark{1,2}, Joshua A.\ Pepper\altaffilmark{3,1}, Ryan J. Oelkers\altaffilmark{1}, Martin Paegert\altaffilmark{4,1}, Nathan De Lee\altaffilmark{5,1}, Roberto Sanchis-Ojeda\altaffilmark{6}}
\altaffiltext{1}{Vanderbilt University, Nashville, TN 37235; keivan.stassun@vanderbilt.edu}
\altaffiltext{2}{Fisk University, Nashville, TN 37208}
\altaffiltext{3}{Lehigh University, Bethlehem, PA 18015}
\altaffiltext{4}{Harvard-Smithsonian Center for Astrophysics, Cambridge, MA 02138}
\altaffiltext{5}{Northern Kentucky University, Highland Heights, KY 41099}
\altaffiltext{6}{Massachusetts Institute of Technology, Cambridge, MA 02139}
\def\teff{$T_{\rm eff}$}

\begin{document}

\begin{abstract}
We provide a catalog of stellar properties for stars observed by the {\it Kepler} follow-on mission, K2. We base the catalog on a cross-match between the K2 Campaign target lists and the current working version of the NASA {\it TESS} target catalog. The resulting K2-TESS Stellar Properties Catalog includes value-added information from the TESS target catalog, including stellar colors, proper motions, effective temperatures, estimated luminosity class (dwarf/subgiant versus giant) based on  reduced-proper-motion, and many other properties via cross-matches to other all-sky catalogs. Also included is the Guest Observer program identification number(s) associated with each K2 target. The K2-TESS Stellar Properties Catalog is available to the community as a freely accessible data portal on the Filtergraph system at: 
\url[http://filtergraph.vanderbilt.edu/tess_k2campaigns]{\texttt{http://filtergraph.vanderbilt.edu/tess\char`_k2campaigns}}. 
\end{abstract}

%\maketitle

\section{Introduction: K2 and the Need for a Stellar Properties Catalog}

The {\it Kepler} telescope was launched in March 2009 and delivered photometry for $\approx$200,000 stars, all located in the same field between Cygnus and Lyra \citep{borucki2010,burke2014}. The main {\it Kepler} mission operated for about 4 years until after a second reaction wheel failed, such that the telescope could no longer be pointed at the original field with sufficient pointing stability. A new mission concept, {\it K2}, was then initiated \citep{howell2014}, in which the telescope is pointed at the ecliptic plane with a new field observed along the ecliptic every three months. {\it K2} should permit discovery of small transiting exoplanets orbiting bright stars, as initial tests show that {\it K2} is capable of photometry with a precision of 20--50 parts per million for thousands of stars in each field \citep[e.g.,][]{vanderburg2014}. 

%Continuing the trend started during the last years of the {\it Kepler} mission, 
It is expected that the community will play a major role on the success of the {\it K2} mission. 
%Nevertheless, dozens of white papers were written by the community in order to select the new mission concept. 
In particular, the target stars to be observed in each {\it K2} field---or Campaign---are selected from community proposals, and for the first eleven fields (Campaigns 0--10) the total number of targets has ranged from 7,000 to 40,000 stars. The brightest and the coolest dwarfs represent two of the most requested types of targets, together with very massive stars, known eclipsing binaries, and stars in open clusters. 
With the first {\it K2} data releases, the {\it Kepler} team has provided coordinates and Kepler magnitudes for all selected stars, but the provision of stellar properties is left as a community endeavor. With this document, we describe an online data portal that aims to provide estimated stellar properties of a large fraction of stars in each {\it K2} Campaign, based on a work-in-progress target catalog that is being developed by the NASA {\it TESS}\footnote{Transiting Exoplanet Survey Satellite (TESS) information is available at: \url{http://tess.gsfc.nasa.gov}.} science team. Our hope is that this resource will serve the community for optimizing the science return of the {\it K2} mission, such as organizing followup observations of {\it K2} targets of interest, for a variety of scientific investigations.

\section{The TESS Target Catalog\label{sec:tess}} 
The Transiting Exoplanet Survey Satellite \citep[TESS;][]{ricker2014}
has been selected by NASA for launch in 2017 as an Astrophysics Explorer mission to
search for planets transiting bright and nearby
stars. 
During its two-year mission, TESS will 
monitor at $\sim$1-minute cadence at least 200\,000 main-sequence dwarf stars with $I_C \approx$ 4--13 to search for planetary transits.

To optimize the TESS target stars for planet detection,
the TESS Science Office's target selection working group (TSWG) is developing a catalog of bright dwarf stars across the sky, from which a final target list for TESS can be drawn based on in-flight observation constraints yet to be determined.  The basic consideration is to assemble a list of dwarf stars all over the sky in the effective temperature (\teff) range of interest to TESS, bright enough for TESS to observe, and taking extra steps to include the scientifically valuable M-dwarfs.  The overall approach is to first combine several all-sky star catalogs to serve as the basis for the target catalog, and then augment that with smaller valuable catalogs.  We then apply cuts to select stars of the desired ranges in apparent magnitude and spectral type, and to eliminate evolved stars.

We use the 2MASS point source catalog as the starting point, since it is the one catalog that exists across the full range of magnitudes for TESS targets for the entire sky.  
%The 2MASS catalog aims to include JHK colors for all its stars.  
We cross-match the 36,597,875 2MASS stars having $J < 13$ against the NOMAD, Tycho2, Hipparcos, APASS, and UCAC4 catalogs. The cross-match is done based primarily on position (1 arcsec tolerance). We also include smaller catalogs of nearby high-proper-motion M-dwarfs. That combined dataset is the Augmented TESS Target Catalog (ATTC), which at the time of this writing contains 
35.8 million stars.

% Old Effective Temperatures
%We use color-\teff\ relations to derive \teff\ from the available optical and infrared colors.  
%Stars with \teff\ $>6650$~K (earlier than F5) are eliminated as too hot. We also extrapolate I-band magnitudes from the JHK colors, and stars fainter than I=13 are eliminated, along with stars between 12 < I < 13 that are hotter than 3840K, so as to keep the set of cool (M0 and later) faint stars. Although TESS is nominally planning to not investigate stars cooler than M5, we currently keep any such stars in the catalog.
%There are a variety of empirical color-\teff\ relations available. Currently we use \teff\ relations from \citet{casagrande2008,casagrande2010}. We calculate \teff\ for each target star in multiple ways and adopt as the best estimate the value with the smallest formal error. The scheme adopted for calculating \teff\ is graphically summarized in Figure~\ref{fig:teff}.

We have developed a procedure to estimate \teff\ based on empirical relationships of stellar $V-K$ color. 
%The procedure is summarized visually in Figure \ref{fig:teff_procedure}. 
The procedure is as follows. 

The color--temperature relation for dwarfs as a function of $V-K_S$ is in two pieces:
\begin{enumerate}
\item For $V-K_S$ in the range $[-0.10, 5.05]$ and [Fe/H] in the range $[-0.9, +0.4]$, we use the AFGKM relation from \citet{huang2015}:
\begin{align*}
X &= V-K_S \, {\rm (de-reddened)} \\
Y &= {\rm [Fe/H]} \\
\theta &= 0.54042 + 0.23676X - 0.00796X^2 - 0.03798XY + 0.05413Y - 0.00448Y^2 \\
T_{\rm eff} &= 5040/\theta
\end{align*}
The scatter of this calibration is 2\% in \teff, which should be added in quadrature to whatever errors come from photometric uncertainties propagated through the above equation.

\item For redder $V-K_S$ values in the range $[5.05, 8.468]$, use the relation from \citet{casagrande2008}, shifted by +205.26 K to meet with the one above at $V-K_S = 5.05$:
\begin{align*}
X &= V-K_S \, {\rm (de-reddened)} \\
\theta &= -0.4809 + 0.8009X - 0.1039X^2 + 0.0056X^3 \\
T_{\rm eff} &= 5040/\theta + 205.26
\end{align*}
The scatter of this calibration is only 19 K, according to the paper, which should be added in quadrature to whatever errors come from photometric uncertainties propagated through the above equation. Note that this Casagrande relation does not include metallicity terms, but those authors claim that the dependence on metallicity is weak for M stars with \teff $>2800$ K.
\end{enumerate}

The above expressions provide a continuous color-temperature relation from 2444 K to 9755 K. For stars with $V-K_S$ outside of the ranges of validity, we report \teff\ = NULL. 
%If the (spectroscopic) metallicity is known but is outside of the range of validity, this is probably not an issue because we will then most likely also have a spectroscopic temperature, so there will be no need to apply these relations.

Finally, for completeness, we provide a similar color-temperature relation for red giants,
%I don't know if we need a relation for giants or not, but here it is anyway 
also taken from \citet{huang2015}:
\begin{align*}
X &= V-K_S \, {\rm (de-reddened)} \\
Y &= \rm {[Fe/H]} \\
\theta &= 0.46447 + 0.30156X - 0.01918X^2 - 0.02526XY + 0.06132Y - 0.04036Y^2 \\
T_{\rm eff} &= 5040/\theta
\end{align*}
which is valid for $V-K_S$ in the range $[1.99, 6.09]$ and [Fe/H] in the range $[-0.6, +0.3]$. The scatter in \teff\ is 1.7\%.

Note that the above relations are all based on $V-K_S$ color. We have a $K_S$ magnitude for nearly every object since 2MASS is the base catalog for the entire TIC. If a $V$ mag is available from APASS, we adopt it (for stars with $V>10$), otherwise we adopt $V_T$ from Tycho, or the $V$ magnitude from Hipparcos. Finally, if none of the above are available, adopt the UCAC $V$ unless the $V$ from UCAC is flagged as unreliable. In all cases, we convert the adopted $V$ to Johnson $V$ using standard conversion relations. 

\begin{figure}[ht]
\includegraphics[width=\linewidth]{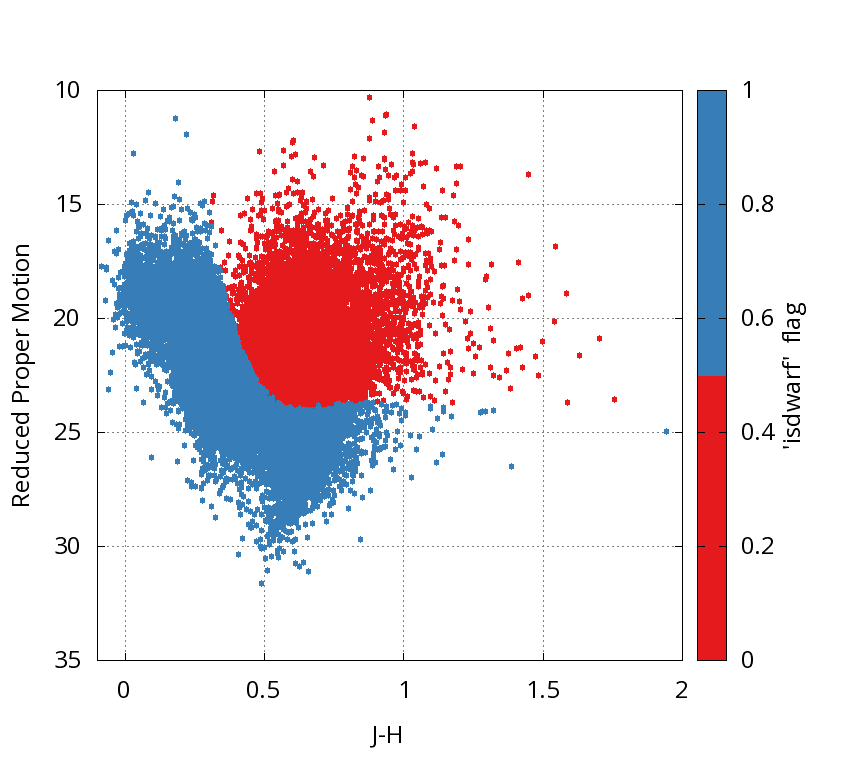}
\caption{We adopt the reduced-proper-motion diagram of Collier Cameron et al.\ (2007) to separate stars into likely giants (red) versus likely dwarfs (blue). Note that subgiants are generally mixed in with the putative dwarfs. 
\label{fig:rpm}}
\end{figure}

%Because transit detectability depends on the radius of the host star, we wish to eliminate evolved stars from the final target list, whose radii make the transit signal too shallow to detect small planets. 
Since spectroscopically determined surface gravities are not available for most stars in the ATTC, we 
%need to use alternate methods for separating giants from dwarfs.  We therefore 
use the reduced proper motion (RPM) statistic, which \citet{collier2007} found to be useful for separating giant stars from dwarfs. Note that the RPM method does not robustly disambiguate subgiants ($3.5 < \log g < 4.1$) from dwarfs ($\log g > 4.2$), so they will be included in the dwarf group. 
However, the method cuts at about $\log(g) = 3.5$. 
For all targets in the ATTC that have recorded proper motions, $\mu$, we compute RPM$_J = J + 5 \log\mu$ (note that this differs by an offset of 5 from the usual definition of RPM$_J = J + 5 \log\mu$ + 5). According to this method, stars with RPM$_J$ less than an empirically-determined cut in RPM$_J$ vs.\ $J-H$ parameter space are taken to be non-giants, i.e.\ either dwarfs or subgiants. We conservatively flag stars that are within 2$\sigma$ of the RPM$_J$ threshold as possible giants. Figure~\ref{fig:rpm} illustrates the separation of K2 stars into giants and dwarfs/subgiants according to the RPM$_J$ cut.

% Martins comment: Do NOT use precise numbers, they will change along the road. The application of those cuts proceeds as follows (see flowchart below).  From the 22,176,076 stars in the ATTC, these steps eliminate 10,572,510 stars with I > 13, another 918,116 that are too red (J-K > 2.0) to derive a reliable I-mag, and 58,063 that are too blue (J-K < -0.5) to derive a reliable I-mag.  Of the remaining 10,626,997 stars with valid I < 13, there are 5,718,452 eliminated as likely giants via the RPMJ cut, and then 340,227 of the remaining stars are eliminated as too hot.  That leaves 4,568,708 stars with I < 13, Teff < 6650K, and are likely dwarfs according to the RPMJ cut.   At that point, 2,216,074 stars that are between 12 < I < 13 and are hotter than 3840K (spectral type M0) are eliminated.  The final set of stars comprises the TESS Dwarf Catalog (TDC), with 2,352,244 members.

\section{The K2-TESS Stellar Properties Catalog}

We have cross-matched the ATTC (see Sec.~\ref{sec:tess}) against the stars observed in all K2 Campaigns\footnote{\label{foot}K2 Campaign target lists obtained from: \url{http://keplerscience.arc.nasa.gov/K2/Fields.shtml}.} (as of this writing this includes Campaigns 0--10).  Since the K2 target stars have mostly been drawn from the EPIC catalog\footnote{The K2 Ecliptic Plane Input Catalog (EPIC) available at: \url{https://archive.stsci.edu/k2}.}, and since the Campaign 0 Engineering Run target list$^{\ref{foot}}$ included target coordinates that were not specified with the precision of the EPIC coordinates, we matched the ATTC directly to EPIC through RA and Dec coordinates (1 arcsec tolerance).  We then select from that overall cross-match the observed K2 campaign stars by their EPIC IDs.  We do not include in the released K2-TESS catalog $\sim 29,000$ stars from K2 Campaigns 0--10 that do not have EPIC coordinates (these are mostly custom apertures for open clusters and solar system objects). The current cross-matched catalog for Campaigns 0--10 includes 117,521 stars with $J<13$.

In addition, while not part of the planned TESS Target Catalog, we have also matched the K2 target stars against all fainter 2MASS and UCAC4 stars in order to provide an additional cross-matched ``faint extended" K2-TESS catalog for $J>13$. We caution that this faint extended catalog is provided as-is. In particular, the performance of the RPM giant/dwarf separation, and of the color-\teff\ relations, have not been as carefully vetted for these fainter stars.

%\subsection{Additions and Improvements}

%\mapins{
Based on the distribution of stars in the $(V-K) - (J-H)$ plane reported in \citet{bessel88}, we developed a method to assign dereddened \teff\ to most of our target stars and report these values in the database. 
%If we have a value for $E(B-V)$ from the Schlegel Dust maps \citep{schlegel98} we derive a dereddening vector for the color plane and stop if we either reach the maximum value given by the dust map or intersect with the location of stars on the color-color plane. For stars outside of the Schlegel maps we assume a fiducial length of the de-reddening vector and stop if we hit the location or where we get closest to it. Depending on the starting location up to 4 solutions are possible. The V-K color of each solution is base for the de-reddened effective temperature. De-reddened (V-K) differences often lie without the validity range of the Casagrande relations, the stars are too hot.

%In order to deal with hot stars we added a relation from \citet{boyajian13}, pushing our maximum effective temperature from 7378 to 9458 K. We added this relation for de-reddened and normal temperatures. While hot stars are of no particular interest for TESS, they are of interest for many users of our portal.

%\mapins{We cross-matched the improved catalog with stellar characteristics from APOGEE 1 \citep{alam2015}, RAVE DR4 \citep{kordopatis13} and LAMOST DR1 \citep{luo15}. For the 63146 stars from K2 campaigns 0 to 5 we find 785 stars with stellar characterristics from APOGEE 1, 2637 stars in RAVE and 2354 A- to K- stars from LAMOST and 672 M-stars which do not have temperatures, but just an M-subclass in LAMOST.}

We also cross-matched the catalog with stellar characteristics from APOGEE-1 \citep{alam2015}, RAVE DR4 \citep{kordopatis13}, and LAMOST DR1 \citep{luo15}. For the 192,884 stars from K2 campaigns 0 to 10 we find 1592 stars with stellar characteristics from Apogee-1, 5336 stars in Rave, 6262 A-K stars from LAMOST, and 1167 M-dwarfs in LAMOST-M (these do not have \teff\ reported in the LAMOST catalog).

\begin{figure}[ht]
\includegraphics[width=\textwidth]{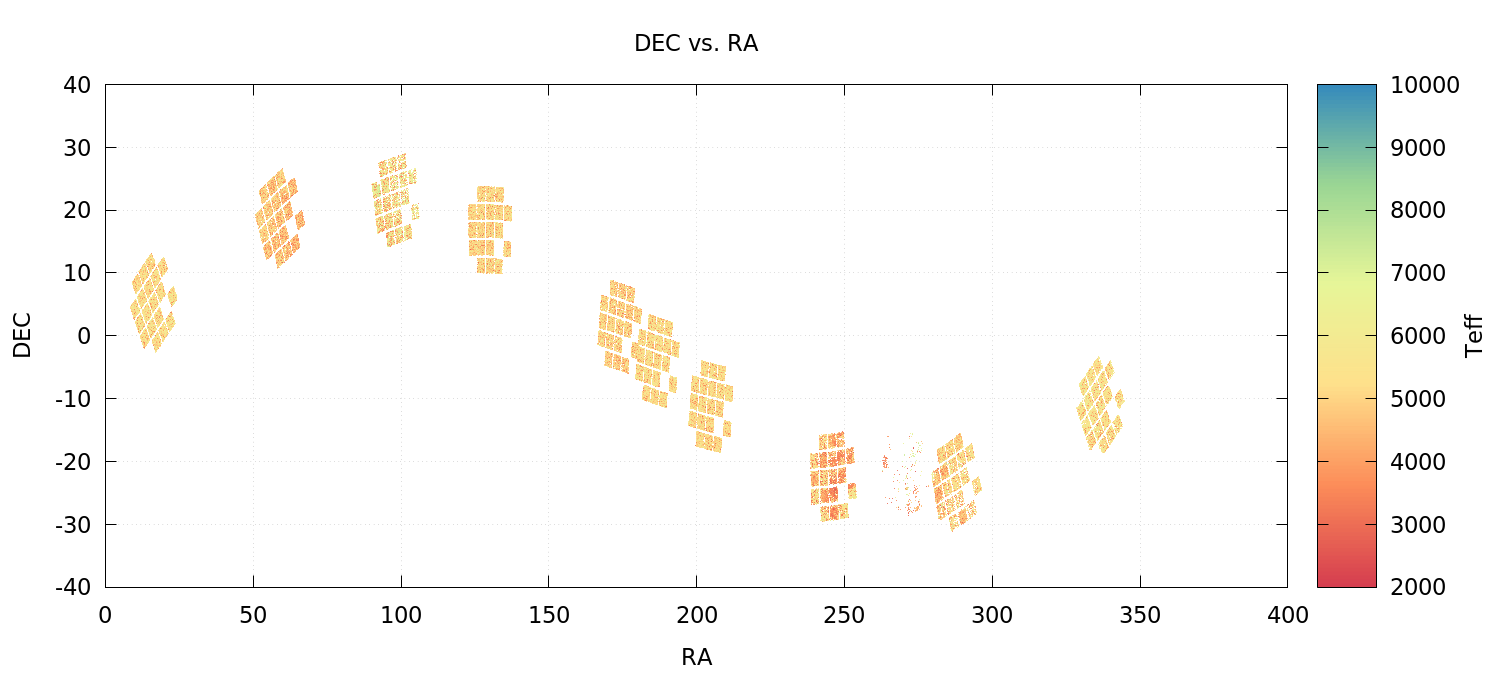} \\
\includegraphics[width=\textwidth]{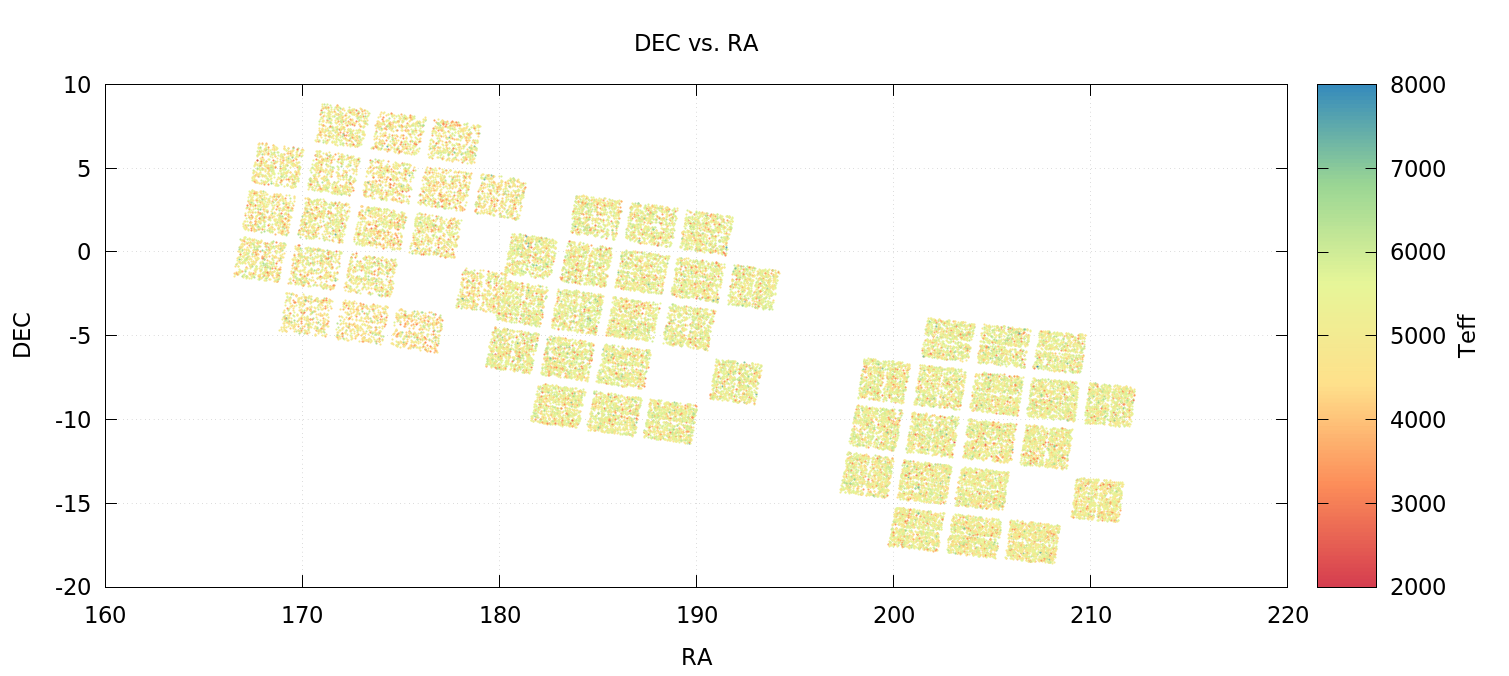}
\caption{Example map displays of all stars with $J<13$ from the K2-TESS catalog on the Filtergraph data portal system. (Top) Full map showing Campaigns 0--10. (Bottom) Zoom of campaigns 1, 10 and 6.
\label{fig:portal}}
\end{figure}

The K2-TESS Stellar Properties Catalog for $J<13$ and the faint extended K2-TESS catalog for $J>13$ are available through the Filtergraph data portal system (Burger et al.\ 2013) at a dedicated URL: 
\url[http://filtergraph.vanderbilt.edu/tess_k2campaigns]{\texttt{http://filtergraph.vanderbilt.edu/tess\char`_k2campaigns}}. Figure~\ref{fig:portal} gives an example map display utilizing the portal, and Table~\ref{tab:fields} gives a listing of the data fields included in the catalog.

%The full K2-TESS catalog includes the following fields: 
%{\bf [Nathan, add a table giving field names and one-line description of each.]}
\begin{table}[!ht]
\centering
\begin{tabular}{c|l}
{\bf Field name} & {\bf Description} \\
\hline
{\tt tmname} & Object identifier from the 2MASS catalog \\
{\tt ra, dec} & Right ascension and declination from 2MASS catalog \\
{\tt glong, glat} & Galactic longitude and latitude from RA and Dec. \\
{\tt ucacname} & UCAC4 catalog identifier \\
{\tt tycname} & Tycho catalog identifier \\
{\tt hipno} & Hipparcos catalog identifier \\
{\tt k2name} & {\it K2}/EPIC target catalog identifier \\
{\tt k2camp} & {\it K2} Campaign number during which target was observed \\
{\tt J, H, K} & Apparent $JHK$ magnitudes from 2MASS \\
{\tt V} & Apparent $V$ magnitude from {\tt Vsrc} \\
{\tt Vsrc} & Catalog source of {\tt V} magnitude \\
{\tt Verr} & Reported error on {\tt V} magnitude from {\tt Vsrc} \\
{\tt pmra, pmdec} & Proper motions in RA and Dec from UCAC4 catalog \\
{\tt isdwarf} & Flag indicating 1 for likely dwarf/subgiant, 0 for likely giant, based on RPM$_J$ criterion \\
{\tt teff} & Estimated \teff\ based on color-\teff\ relation using photometry from {\tt teffsrc} \\
{\tt teffsrc} & Catalog source of photometry used to derive \teff\ ({\tt teff}) \\
{\tt tefferr} & Propagated uncertainty on \teff\ ({\tt teff}) \\
{\tt drteff$_{i}$} & Up to 4 solutions for de-reddened \teff \\
{\tt kepmag} & Apparent magnitude in the {\it Kepler} bandpass, taken from EPIC catalog \\
{\tt investids} & {\it K2} GO ID numbers associated with target (multiple IDs separated by `$|$') \\
{\tt apgteff, apgtefferr} & \teff and error from APOGEE 1\\
{\tt apglogg, apgloggerr} & $\log(g)$ and error from APOGEE 1\\
{\tt apgmh, apgmherr} & metallicity and error from APOGEE 1\\
{\tt raveteff, ravetefferr} & \teff and error from RAVE DR4 (stellar characteristics)\\
{\tt ravelogg, raveloggerr} & $\log(g)$ and error from RAVE DR4 (stellar characteristics)\\
{\tt ravemet, ravemeterr} & metallicity and error from RAVE DR4 (stellar characteristics)\\
{\tt raveteffv} & \teff from RAVE DR7 (radial velocity pipeline)\\
{\tt raveloggv} & $\log(g)$ and error from RAVE DR4 (radial velocity pipeline)\\
{\tt lmakteff, lmaktefferr} & \teff and error from LAMOST DR1\\
{\tt lmaklogg, lmakloggerr} & $\log(g)$ and error from LAMOST DR1\\
{\tt lmakfeh, lmakfeherr} & metallicity and error from LAMOST DR1\\
{\tt lmmsubclass} & subclass for M-stars from LAMOST DR1\\
\end{tabular}

\caption{{\it Description of fields in the K2-TESS Stellar Properties catalog and K2-TESS faint extended catalog.}
\label{tab:fields}}
\end{table}

The K2-TESS Stellar Properties catalog is being provided as a service to the community. We intend to regularly update the catalog on the Filtergraph data portal as additional K2 Campaign targets are observed. The data portal website includes fair-use terms and contact information.
We expect to have a significantly improved portal in the next two months based on the new TESS Input Catalog matches to Tycho-2, the inclusion of bright star $V$ magnitudes, and improved calculations of \teff\ and dereddened \teff.

\end{document}